\documentclass[aps,prd,10pt,twocolumn,nofootinbib,floatfix,
  noshowpacs,preprintnumbers]{revtex4-1}
\usepackage{bm}
\usepackage{epsfig}
\usepackage{amsmath}
\usepackage{color}
\usepackage[dvipsnames]{xcolor}
\usepackage[colorlinks=true,breaklinks=true]{hyperref}
\hypersetup{allcolors=[rgb]{0.0 0.0 0.6},linkcolor=[rgb]{0.75 0.05 0.05}}
\usepackage[normalem]{ulem}

\newcommand{\exclude}[1]{}

\def\w {\omega}

\begin{document}

\title{Supernova Neutrinos: flavour conversion mechanisms and new physics scenarios}

\author{Manibrata Sen}
\email{manibrata@mpi-hd.mpg.de}
\affiliation{Max-Planck-Institut für Kernphysik,
Saupfercheckweg 1, 69117 Heidelberg, Germany.
}

\date\today

\begin{abstract}
A core-collapse supernova (SN) releases almost all of its energy in the form of neutrinos, which provide a unique opportunity to probe the working machinery of a SN. These sites are prone to neutrino-neutrino refractive effects, which can lead to fascinating collective flavour oscillations among neutrinos. This causes rapid neutrino flavour conversions deep inside the SN even for suppressed mixing angles, with intriguing consequences for the explosion mechanism as well as nucleosynthesis. We review the physics of collective oscillations of neutrinos- both slow and fast, along with the well-known resonant flavour conversion effects, and discuss the current state-of-the-art of the field. Furthermore, we discuss how neutrinos from a SN can be used to probe novel particle physics properties, extreme values of which are otherwise inaccessible in laboratories.
\end{abstract}

\maketitle

\section{Introduction}
\label{sec:Introduction}
The observation of neutrinos originating from the supernova SN1987A stands as a pioneering moment in the realm of multi-messenger physics. This pivotal event involved the detection of neutrino signals several hours before the emergence of visible light. Despite the modest quantity of data, encompassing roughly 30 neutrino events recorded by Kamiokande-II (KII)~\cite{Hirata:1987hu,Hirata:1988ad}, IMB~\cite{Bionta:1987qt,Bratton:1988ww}, and Baksan~\cite{Alekseev:1988gp}, it has garnered considerable attention within the fields of particle and astroparticle physics, serving as a valuable resource for exploring fundamental questions in the discipline.

The working machinery of a core-collapse supernovae (SN) can be briefly described as follows (for detailed reviews, see Refs.\,\cite{Bethe:1990mw,Mezzacappa:2005ju,Raffelt:1996wa,Kotake:2005zn,Woosley:2006ie,Janka:2006fh,Janka:2012wk,Mirizzi:2015eza,Burrows:2020qrp}). At the end of nuclear burning, a massive star (more than $8\,M_\odot$) ends up with an iron core with concentric shells of lighter elements above it. As the iron core grows in mass due to accretion and crosses the Chandrasekhar mass limit, electron degeneracy pressure fails to support it, and a collapse is triggered. The collapse proceeds until nuclear densities are reached, which is when the equation of state becomes stiff and resists further infall, resulting in a core bounce. This launches an energetic shockwave which travels outwards from the core, which is responsible for exploding the star.

Simulations suggest that this shockwave is not powerful enough to blow up the star, rather it loses energy and stalls at a radius of around 100-200 km from the core~\cite{1934PNAS...20..254B,Brown:1982cpw}. As the shock stalls, the compact protoneutron star at the centre begins to grow by accretion of infalling stellar material.  In this entire process, a lot of neutrinos are released, which start free-streaming from the neutrinosphere- the surface of last scattering- at a radius of around 30 km from the core~\cite{Burrows:1990ts}. It is widely believed that the emitted neutrinos deposit their energy in the region behind the shock and re-energise the shockwave causing a successful explosion, also known as the delayed neutrino-heating mechanism of explosion~\cite{Colgate:1966ax,Bethe:1984ux}. 
This scenario is highly sensitive to the neutrino flavour information and hence requires a precise understanding of neutrino flavour oscillations. The question of SN explosion also depended on the SN geometry used in simulations. Initial studies focussed on simplistic spherically symmetric (1D) SN models seldom ended in an explosion, however, we are now in an era of 3D simulations, which can capture the multidimensional turbulent convection in the SN envelope in detail. This has led to a better understanding of the role played by convection as well as the breaking of spherical symmetry in SN explosions~\cite{Burrows:2020qrp}.

It is widely acknowledged that investigating neutrinos emitted during the core collapse of a galactic SN offers a unique opportunity to delve into the dynamics of these explosive phenomena. These neutrinos originate from deep inside the supernova and traverse the entire matter envelope before reaching Earth. The composition of neutrino flavours is exquisitely sensitive to the fermion density they encounter during their journey.

Before the turn of the millennium, prevailing scientific belief centred on resonant flavour conversions within the SN, driven by the Mikheyev-Smirnov-Wolfenstein (MSW) effect~\cite{Mikheev:1986gs,Wolfenstein:1977ue,Kuo:1988pn,Dighe:1999bi}. This perspective, however, overlooked a fundamental factor: the impact of neutrino self-interactions, a critical element governing neutrino flavour dynamics within a dense medium, as discussed by Pantaleone in two classic papers~\cite{Pantaleone:1992eq,Pantaleone:1994ns}. In regions near the neutrinosphere, where neutrino density is substantial, forward scattering of neutrinos on a background of other neutrinos prevails as the dominant factor in the propagation Hamiltonian~\cite{Kostelecky:1994dt,Samuel:1995ri,Duan:2005cp,Duan:2006an,Duan:2006jv}. These self-interactions manifest as a potential linked to the background neutrino density, denoted by $n_\nu$. Furthermore, neutral current interactions among neutrinos of different flavours, characterised by their indifference to flavour, also contribute off-diagonal components to this forward-scattering potential. Consequently, neutrinos were found to undergo collective flavour transformation, unveiling a diverse range of phenomena~\cite{Duan:2005cp,Duan:2006an,Duan:2006jv}.

Until the first decade of the millennium, the prevailing belief was that these collective oscillations, termed as bipolar flavour conversions, would grow with a rate roughly linked to $\sqrt{(\Delta m^2) n_\nu}$, where $\Delta m^2$ represents the neutrino mass-squared difference~\cite{Duan:2006an, Hannestad:2006nj,Duan:2007mv}. Simplistic toy models indicated that such flavour conversions occurred in the vicinity of the stalled shockwave, at approximately 150-200 km from the core. This hinted at their potential involvement in facilitating a shockwave-driven explosion. However, as research progressed, it became evident that moving beyond the initial single-angle approximation, which assumed uniform emission angles for neutrinos of all flavours from a specific neutrinosphere, revealed these flavour conversions to happen at radii too distant from the core to significantly influence the SN explosion mechanism (for a detailed review, see \cite{Duan:2010bg}).

More recently, researchers have recognised an entirely distinct phenomenon known as ``fast" flavour conversions, which can manifest in the depths of SNe and neutron star mergers~\cite{Sawyer:2005jk,Sawyer:2008zs,Sawyer:2015dsa,Chakraborty:2016lct,Dasgupta:2016dbv}. This stems from distinct angular  distributions of neutrinos and antineutrinos across different flavours. It can thrive in regions inaccessible to other flavour transformation mechanisms. Even for massless neutrinos, small perturbations can amplify with a rate proportional to $n_\nu$, and the rate remains almost independent of neutrino masses and mass hierarchy (see~\cite{Dasgupta:2017oko,Shalgar:2020xns} for a detailed study of the dependence). It is believed that one of the most likely outcomes of such rapid flavour conversions is flavour depolarization, which can lead to equipartition of neutrino flavours inside a SN.
This can have drastic consequences for the explosion mechanism, as well as nucleosynthesis.  From a neutrino transport simulation standpoint,  such a flavour equipartition implies that neutrino oscillations will not play an important role in the supernova dynamics, hence this can be a huge gain as this partially obliterates the need to include neutrino oscillations into a transport code, which is computationally very expensive. These developments have led to intense research work to assess the effect of these flavour conversions on the SN explosion
mechanism (see~\cite{Tamborra:2020cul,Richers:2022zug} for a recent review).  However, we are still far from having a complete picture of the potential impact of these collective oscillations.

The neutrino flux from a SN can be utilised to facilitate a SN as a unique laboratory for scientific exploration~\cite{Raffelt:1996wa,Horiuchi:2017sku}. The extreme conditions prevailing within these objects provide a rare opportunity to impose stringent constraints on neutrino properties, which often reach extreme values not achievable in terrestrial laboratories. This is exemplified by the limited number of neutrino events detected from SN1987A, which have enabled the imposition of robust constraints on physics beyond the Standard Model. Remarkably, even hypothetical ``feebly'' interacting particles can be effectively constrained through luminosity arguments derived from supernova observations.

The few neutrino events observed from SN1987A were sufficient to confirm our basic understanding of the working machinery of a core-collapse SN~\cite{Fiorillo:2023frv,Li:2023ulf}. On average, galactic SNe are expected to occur once every 30-50 years, which implies that we need to be prepared for another one anytime in the next few decades. We are in a period of enormous advancement in supernova neutrino detection experiments. Among the currently running experiments, Super-Kamiokande (Japan), with the addition of Gd~\cite{Super-Kamiokande:2021the}, can detect the highest number of SN neutrino events. Among the liquid scintillator detectors, KamLAND (Japan)~\cite{KamLAND:2022sqb} offer the best bets for detection of these neutrinos. The IceCube detector at the South Pole does not have enough sensitivity to detect individual events, but it can observe a SN burst as a correlated rise of background noise~\cite{2011JPhCS.309a2029K}. Experiments like the Deep Underground Neutrino Experiment (USA)~\cite{DUNE:2020ypp}, Hyper-Kamiokande (Japan)~\cite{Hyper-Kamiokande:2018ofw} and Jiangmen Underground Neutrino Observatory (China)~\cite{JUNO:2015zny} offer immense potential in terms of deducing the unknowns of SN neutrino spectra. With improving technologies, even coherent elastic neutrino-nucleus scattering experiments can play a major role in SN neutrino detection~\cite{Pattavina:2020cqc}. For a more detailed discussion on the aspects of detection, please refer to~\cite{Mirizzi:2015eza} and references therein

In this brief review, we summarise the current status of the field, focusing on neutrino oscillation physics, as well as the different probes of new physics that can be achieved from the observation of supernova neutrinos. We would also like to point interested readers to a list of other comprehensive reviews on various aspects of this topic~\cite{Mirizzi:2015eza,Chakraborty:2015tfa,Horiuchi:2017sku,Fuller:2022nbn,Volpe:2023met,Duan:2010bg, Tamborra:2020cul,Richers:2022zug,Capozzi:2022slf}.
This article is organised as follows. In Sec.\,\ref{sec:SNprop}, we discuss the basic picture of neutrino emission from a SN, followed by a discussion of the governing equations of motion in Sec.\,\ref{sec:eom}. Then we expand on the aspect of neutrino flavour propagation inside a SN, focussing on recent developments in collective oscillations, in Sec.\,\ref{sec:flavconv}. Finally, we describe the various kinds of new physics scenarios which can be tested from the observation of neutrinos from a SN in Sec.\,\ref{sec:newphys} before concluding in Sec.\,\ref{sec:conc}.

\section{Supernova neutrino: emission phases and spectra}
\label{sec:SNprop}
It is anticipated that approximately $10^{58}$ neutrinos are emitted from a SN explosion. This emission occurs over a timescale of approximately 10 seconds, which corresponds to the typical diffusion timescale of neutrinos within the core. The emission of neutrinos is expected to occur in four distinct phases~\cite{Colgate:1966ax,Bethe:1984ux,1984ApJ...283..848B,Raffelt:1996wa}:

\begin{enumerate}
    \item During the initial collapse phase, preceding the core bounce, a small amount of electron neutrinos is primarily emitted due to processes like beta decay and electron capture.

    \item Following the core bounce, approximately 25 milliseconds later, there is a rapid increase in the flux of electron neutrinos known as the ``neutronization burst" phase. During this phase, small amounts of electron antineutrinos and other neutrino flavours are also emitted, although their contribution is negligible compared to the flux of electron neutrinos.

    \item  Subsequently, the ``accretion phase" ensues, lasting for approximately a few hundred milliseconds. As the shockwave stalls, matter accumulates onto the core over several hundred milliseconds. During this phase, the star undergoes cooling by emitting neutrinos and antineutrinos of all flavours. Typically, there is an excess of electron neutrinos compared to other species, with a discernible hierarchy in average energies: $\langle E_{\nu_{\mu,\tau}}\rangle > \langle E_{\bar{\nu}_e} \rangle > \langle E_{\nu_{e}}\rangle $.

    \item  Finally, the ``Kelvin-Helmholtz cooling phase" begins, wherein the proto-neutron star at the core cools down by emitting neutrinos over a period of around ten seconds. In this phase, the hierarchy of average energy among neutrino flavours is observed to be less pronounced.
\end{enumerate}
SN simulations fit the neutrino spectra with a ``quasi-thermal'' spectra, known in the literature as the ``alpha-fit''~\cite{Keil:2002in,Tamborra:2012ac}:
 \begin{equation}\label{spectra_ch1}
  f_{\nu_\alpha}(E)=\frac{1}{\langle E_\nu \rangle}\frac{(\alpha+1)^{(\alpha+1)}}{\Gamma (\alpha+1)}\left(\frac{E}{\langle E_\nu \rangle}\right)^{\alpha}{\rm e}^{-(\alpha+1) \frac{E}{\langle E_\nu \rangle}}\,.
 \end{equation}
Here, $\Gamma(z)$ is the Euler gamma function, and $\alpha$ is the pinching parameter which gives the deviation from a thermal spectrum, given by
\begin{equation}
 \frac{1}{1+\alpha}=\frac{\langle E^2\rangle-\langle E\rangle^2}{\langle E\rangle^2}\,.
\end{equation}
A value of $\alpha=2$ corresponds to a Maxwell-Boltzmann spectrum, while $\alpha >2$ signifies a pinched spectrum, characterised by suppressed high and low energy tails. Conversely, $\alpha<2$ denotes a broadened spectrum. Commonly used simulations typically yield values of $\alpha$ falling within the range $2\leq\alpha\leq5$ \cite{Keil:2002in}. It is generally expected that for neutrinos, $\langle E_{\nu_e}\rangle \in (10\,{\rm MeV}, 12\,$MeV), for antineutrinos $\langle  E_{\bar{\nu}_e}\rangle\in (12\,{\rm MeV}, 15\,{\rm MeV})$, while for the non-electron flavour neutrinos, we have $\langle E_{\nu_{\mu,\tau}}\rangle \in (15\,{\rm MeV},  18\,{\rm MeV})$, although the exact values depend on the simulation. The observed hierarchy in the average energies of different neutrino flavours primarily stems from variations in their interaction cross-sections. For example, due to larger cross-section and charged-current interaction with neutrons, which are more abundant than protons, $\nu_e$ have a lower average energy than $\bar{\nu}_e$.  On the other hand, since $\nu_{\mu,\tau}$ interacts less than $\nu_e$ and $\bar{\nu}_e$, their average energy is expected to be higher, and hence the hierarchy.
While initial studies ignored the presence of muons and taus in SN simulations, and assumed a similar spectra for $\nu_{\mu,\tau}$, more recent simulations have shown that substantial amount of muons can be present due to the high electron chemical potential, and hence cannot be neglected~\cite{Bollig:2017lki}. It has been shown that the addition of muons considerably softens the equation of state of the core and thus faciliates the contraction of the protoneutron star, leading to larger luminosities and average energies for neutrinos.

\section{Neutrino propagation: equations of motion}
\label{sec:eom}
The straightforward approach of applying the Schrödinger equation to address neutrino mixing encounters limitations when dealing with the dynamics of a statistical ensemble of neutrinos engaged in both mixing and scattering within a medium. A more effective approach necessitates the utilization of the density matrix formalism, which allows for the comprehensive consideration of these effects. Within this framework, it becomes possible to depict the mixing of quantum states among neutrinos, accounting for potential coherence loss resulting from actual collisions. In this section, we will outline the equations of motion using an effective two-flavour setup, consisting of $\nu_e$ and $\nu_x$, where $\nu_x$ denotes a linear combination of the $\nu_\mu-\nu_\tau$ flavours. Generalisation to three flavours is straightforward. 

An ensemble of neutrinos and antineutrinos can be characterised by the $2 \times 2$  density
matrices $\varrho_{{\bf p}, {\bf r},t}$ (  $\overline{\varrho}_{{\bf p}, {\bf r},t}$ for antineutrinos)\cite{Raffelt:1991ck, Raffelt:1992uj,Dolgov:1980cq, Barbieri:1990vx, Sigl:1993ctk},
\begin{align}
 \small{
\varrho_{{\bf p}, {\bf r},t}=
\left(\begin{array}{cc} 
\varrho_{ee} &  \varrho_{e x} \\
\varrho_{x e}  & \varrho_{x x}  \\
\end{array}\right)}~, 
\label{eq:rho}
\end{align}
where the diagonal elements of $\varrho_{{\bf p}, {\bf r},t}$  are proportional to the occupation numbers for the corresponding flavours, while the off-diagonal ones give the phase information.
At this point, it is crucial to mention that these entries are essentially expectation values of the bilinears of the neutrino fields. This so-called mean-field approximation neglects any possible entanglement among the neutrinos. We will later discuss the implications of relaxing these assumptions. 


The evolution of the density matrices $\rho_{{\bf p}, {\bf r},t}$ for momentum ${\bf p}$ at position ${\bf r}$ and time $t$ is governed by the equations of motion (EoMs)~\cite{Sigl:1993ctk}
\begin{equation}
\left(\partial_t + {\bf v}_{\bf p} \cdot \nabla_{\bf r}\right) \rho_{{\bf p}, {\bf r},t} 
= -i [\Omega_{{\bf p}, {\bf r},t}, \rho_{{\bf p}, {\bf r},t}] 
\,\, + C[\rho_{{\bf p}, {\bf r},t} ]\,,
\label{eq:eom}
\end{equation}
where the left-hand side accounts for time dependence and spatial drift caused by free-streaming neutrinos. The Hamiltonian consists of three terms, i.e.,  $\Omega_{{\bf p}, {\bf r},t}= \Omega_{\rm vac} + \Omega_{\rm MSW} + \Omega_{\nu \nu}$. These contributions comprise the vacuum term,
\begin{equation}
\label{eq:vac}
    \Omega_{\rm vac}=\frac{\w}{2}\begin{pmatrix}
        -1 & 0\\
        0  & 1
    \end{pmatrix}\,,
\end{equation}
where $\w=\Delta m^2/2\,E$; the Mikheyev-Smirnov-Wolfenstein matter potential term~\cite{Wolfenstein:1977ue,Mikheev:1986gs} 
\begin{equation}
    \Omega_{\rm MSW}=\sqrt{2}G_F n_e \begin{pmatrix}
        1 & 0\\
        0  & 0
    \end{pmatrix}\,,
\end{equation}
where $n_e$ denotes the background electron number density and $G_F$ is the Fermi constant; and the $\nu-\nu$ self-interaction potential~\cite{Pantaleone:1992eq,Pantaleone:1994ns}, 
\begin{equation}
    \Omega_{\nu\nu}=\sqrt{2}G_F\int \frac{d^3{\bf q}}{(2\pi)^3} (1-{\bf v}_{\bf p}\cdot {\bf v}_{\bf q})({\rho}_{{\bf q}, {\bf r},t}-{\bar\rho}_{{\bf q}, {\bf r},t})\,. 
\end{equation}
The self-interaction is usually described in terms of an overall normalisation $\mu=\sqrt{2}G_F n_\nu$, where $n_\nu$ is the net neutrino density. A similar equation exists for antineutrinos with a relative change in sign for ${\Omega}^{\rm vac}$ and ${\Omega}^{\rm ref}$. 
In addition to this, neutrinos can also undergo inelastic number-changing/preserving collisions, which are of order $G_F^2$, and only relevant deep inside the SN. These collisional processes are denoted symbolically through $C[\rho_{{\bf p}, {\bf r},t} ]$.

The solutions to the EoMs govern the flavour evolution of neutrinos inside a SN. 
These equations constitute a complex set of non-linear, interrelated partial differential equations that encompass three spatial dimensions, three momentum dimensions, and one temporal dimension. As such, they pose a formidable challenge for analytical solutions. Even numerical solutions are arduous to obtain for the entire problem, necessitating simplifications. In the next section, we will delve into the dynamics of the flavour evolution of neutrinos that can be gauged by solving these equations under simplifying assumptions and for toy setups.

\section{Neutrino flavour conversions inside a supernova}
\label{sec:flavconv}
In this section, we discuss the various flavour conversion mechanisms relevant to SN neutrinos. We begin with resonant adiabatic MSW flavour conversions, which were initially believed to be the major mechanism of flavour conversions inside a SN. Then we discuss the collective flavour oscillations which were realised at the beginning of the millennium. Finally, we shed some light on fast flavour conversions - a unique type of flavour conversion mechanism operating deep inside the SN.
\subsection{Matter induced resonant flavour conversions}
Resonant flavour conversions of neutrinos arise from coherent forward scattering interactions with background fermions. These interactions primarily involve one power of the Fermi coupling $G_F$, dominating over inelastic and loop processes, which typically entail larger powers of $G_F$~\cite{Wolfenstein:1977ue}. In ordinary matter, which comprises electrons, protons, and neutrons but not muons or taus generally, forward scattering processes lead to charged current (CC) interactions between electron neutrinos and electrons, as well as neutral current (NC) interactions among all neutrino flavours and neutrons, protons, and electrons. In this section, we operate in the limit where the neutrino-neutrino self-interaction term $\Omega_{\rm MSW}\gg\Omega_{\nu \nu}$, which typically happens at larger radii inside a SN, where we observe a decline in neutrino density.

The coherent effects of these processes on neutrinos can be likened to neutrino propagation within an effective potential~\cite{Wolfenstein:1977ue,Mikheev:1986gs,Kuo:1989qe}. The effective potential due to CC interactions with electron neutrinos is expressed as,
\begin{equation}
V_{\rm CC}(r)=\sqrt{2}G_F n_e(r)\,,
\end{equation}
where $n_e(r)$ represents the local electron density in the medium. Similarly, for NC interactions involving neutrinos and nucleons, the effective potential is denoted as,
\begin{equation}
V_{\rm NC}(r)=-\frac{G_F n_n(r)}{\sqrt{2}}\,, 
\end{equation}
where $n_n(r)$ denotes the local nucleon density in the medium. It is important to note that since NC interactions are flavour-blind, this potential remains the same for all neutrino flavours and can therefore be eliminated by a phase rotation. For antineutrinos, both these effective potentials acquire a relative minus sign.

This alters the dispersion relation of neutrinos in a medium. Within a simple two-flavour framework, it can be shown that for a constant matter density, the oscillation probability can reach a maximum when~\cite{Mikheev:1986gs}
\begin{equation}\label{MSW}
\frac{\Delta m^2}{2 E}\,\cos2\vartheta=V_{\rm CC}\,,
\end{equation}
where $\Delta m^2$ is the mass-squared difference between the two mass eigenstates in vacuum,  $E$ is the neutrino energy, and $\vartheta$ is the two-flavour mixing angle. This is the MSW (Mikheyev-Smirnov-Wolfenstein) resonance condition and can lead to a maximal mixing in matter, causing large flavour conditions even when $\vartheta$ is small. For a varying matter density, flavour propagation is more complicated and depends on whether the matter density changes adiabatically or in a non-adiabatic manner~\cite{Kuo:1988pn}.

Numerous investigations have underscored the sensitivity of neutrino spectra to the underlying density profile of a supernova, with a particular emphasis on the MSW resonances~\cite{Kuo:1988pn,Dighe:1999bi}. Within a SN, the mass-squared differences relevant to solar ($\Delta m^2_\odot$) and atmospheric ($\Delta m^2_{\text{atm}}$) neutrinos exhibit distinct resonance behaviours, termed the ``L-resonance'' and ``H-resonance'' respectively. Since $\Delta m^2_\odot>0$, the ``L-resonance'' is known to take place for neutrinos for typical SN matter densities around $\rho_L \approx (10-100)\, \text{g/cc}$. On the other hand, the H-resonance, manifesting at matter densities approximately $\rho_H \approx (10^3-10^4) \, \text{g/cc}$, pertains to atmospheric neutrinos in the normal mass ordering (NMO) and antineutrinos in the inverted mass ordering (IMO), owing to the sign of $\Delta m^2_{\text{atm}}$. Global analyses of neutrino data suggest that both resonances are adiabatic, given the substantial mixing angles~\cite{nufit,Esteban:2016qun}. Extensive investigations have explored these resonant flavour conversions within supernovae, contributing to our understanding of neutrino properties and supernova dynamics~\cite{Dighe:1999bi,1987ApJ...322..795F,Lunardini:2003eh,Dighe:2003be,Dighe:2003jg,Dighe:2003vm,Barger:2005it,Fogli:2002xj}.

The matter effects also offer insights into the propagation of the SN shockwave at later stages~\cite{Schirato:2002tg,Takahashi:2002yj,Lunardini:2003eh,Fogli:2003dw,Fogli:2004ff,Tomas:2004gr,Dasgupta:2005wn,Galais:2009wi}.
Typically, non-adiabaticity in neutrino propagation can surface when abrupt density changes occur at the shockfront, leaving a distinctive mark on flavour transitions. This phenomenon becomes a means to reconstruct the real-time progression of the shockwave. A characteristic manifestation of non-adiabaticity due to shockwave propagation can be observed in the form of energy-dependent dips in neutrino survival probabilities~\cite{Tomas:2004gr}. A few seconds post the core bounce in a supernova, the potential emergence of a reverse shockwave hinges on whether a supersonic neutrino-driven wind intersects with the ejecta. The sensitivity of this phenomenon to the supersonic or subsonic nature of the neutrino-driven wind has been thoroughly explored recently~\cite{Friedland:2020ecy}.

Nevertheless, the story does not end here. Supernova progenitors harbour an array of hydrodynamic instabilities that materialise behind the shockfront, generating turbulence within the ejecta. This turbulence, often present in aspherical supernova simulations, introduces stochastic fluctuations in matter density~\cite{Radice:2017kmj}. These fluctuations, in turn, exert a non-trivial influence on neutrino flavour evolution. Extensive research has been conducted to investigate neutrino flavour evolution in turbulent media, especially in the context of predicting neutrino spectra from future galactic supernovae~~\cite{Fogli:2006xy,friedland2006neutrino,Kneller:2010sc,Borriello:2013tha, Lund:2013uta,Kneller:2014oea,Patton:2014lza,Yang:2015oya,Kneller:2017lqg,Abbar:2020ror,Mukhopadhyay:2023tsc}. The impact of turbulence on neutrino survival probability has been studied assuming power-law fluctuations in the density profile~\cite{Borriello:2013tha}. As the turbulence amplitude increased, a power-law behaviour was observed in the electron neutrino survival probability. This led to an analytical estimate explaining the power-law phenomenon within the framework of the simple turbulence model~\cite{friedland2006neutrino}. Subsequent works employed various techniques, including the rotating wave approximation~\cite{Patton:2014lza,Yang:2015oya,Kneller:2017lqg}, to predict the final survival probability and study the dependence of neutrino flavour transition on turbulence amplitude and spectral index~\cite{Mukhopadhyay:2023tsc}.

Finally, the neutrinos are also expected to undergo matter effects as they traverse through the Earth to the detector. These effects, aptly dubbed as ``Earth-matter'' effects, can be quite relevant for SN neutrino neutrinos~\cite{Lunardini:2000sw,Lunardini:2001pb,Dighe:2003jg,Mirizzi:2006xx,Dasgupta:2008my,Borriello:2012zc}. It was found that the major outcome of this effect is an oscillatory modulation of the (anti)neutrino spectra. This effect increases with the neutrino 
energy and can be larger than $\mathcal{O}(20\%)$ for $E_\nu \gtrsim 20\,{\rm MeV}$. The relevant Earth-matter effects can be gauged by comparing the neutrino spectra at different detectors. However, due to the inherent uncertainty concerning the SN neutrino spectra, this effect is expected to be subdominant, and only relevant in large upcoming detectors. In fact, it was shown that the Earth-matter effects can be invoked to explain the discrepancy in the SN1987A neutrino events detected by Kamiokande and IMB, as well as explaining the lack of any neutrino events beyond 40~MeV~\cite{Lunardini:2000sw}.
Earth-matter effects of SN neutrinos have also been shown to play an important role in Earth tomography with neutrinos~\cite{Hajjar:2023knk}.

\subsection{Slow collective flavour conversions}
In the first decade of the millennium, the impact of the neutrino self-interaction term on flavour propagation started getting investigated. The non-linear nature of the governing equations necessitated the introduction of simplifications. One such simplification was the adoption of the single-angle approximation~\cite{Duan:2005cp,Duan:2006an,Duan:2006jv,Hannestad:2006nj,Duan:2007mv}, where neutrinos and antineutrinos escape from the neutrinosphere with a single angle. This reduces the evolution to one degree of freedom (namely, the line of sight), and hence either temporal or spatial evolution was considered.

\begin{figure}[!t]
    \includegraphics[width=0.8\linewidth]{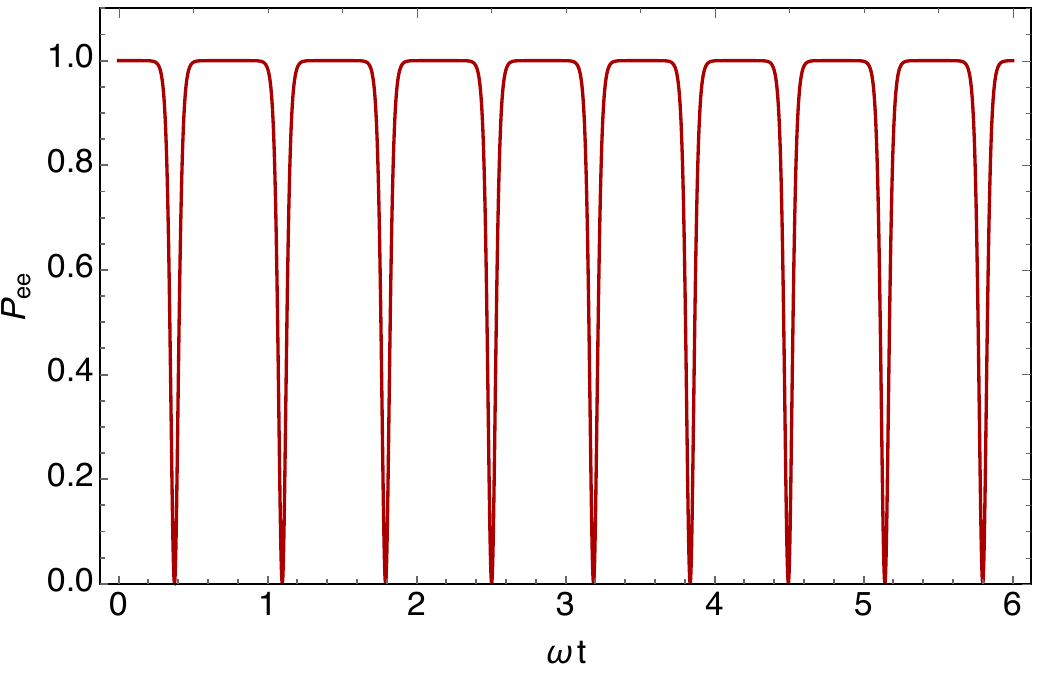}\\ 
    \caption{An example setup of interacting neutrino beams showing bipolar collective oscillations. The survival probability of one of the beams is plotted as a function of time, in units of $\omega$.}
    \label{fig:bipolar}
\end{figure}
Under these approximations, a simple system of interacting neutrino beams was shown to undergo collective effects in flavour evolution. For $\mu \gg \omega$, the system undergoes ``synchronized oscillations'', where neutrinos of different energies oscillate with a common frequency with an amplitude proportional to the suppressed mixing angle (initial phase in Fig.\,\ref{fig:bipolar}). However, as $\mu$ decreases, large amplitude flavour oscillations are seen to happen even with tiny mixing angles. The phenomenon, known as ``bipolar oscillations", involves pair conversions between $\nu_e\bar{\nu}_e \leftrightarrow \nu_x\bar{\nu}_x$. In this scenario, all neutrino energies oscillate with an averaged uniform frequency, approximately proportional to $\sqrt{\omega\mu}$, where $\omega$ has been defined following Eq.\,\ref{eq:vac}. For monochromatic neutrino beams, the system can be mathematically likened to a pendulum in flavour space, akin to the equivalence between ordinary neutrino oscillations in vacuum or matter and the precession of a spin~\cite{Hannestad:2006nj,Duan:2007mv}. Depending on the neutrino mass ordering, the gravitational force acting on this flavour pendulum can be directed either upward or downward, rendering specific flavour configurations unstable, much like an inverted pendulum. Bipolar oscillations involve the pendulum's initial position being in an unstable, inverted state, slightly offset by a small mixing angle, and swinging through the lowest point to the opposite side. 
Just like the time period of an inverted pendulum depends logarithmically on the initial tilting angle from the vertical, the period of the flavour pendulum oscillations also exhibits a similar logarithmic dependence on the vacuum mixing angle.  

The introduction of multi-angle effects also led to the presence of additional instabilities in flavour evolution~\cite{Sawyer:2008zs,Raffelt:2013rqa}. Nevertheless, during the accretion phase, the influence of multi-angle matter effects was demonstrated to suppress collective oscillations~ \cite{Chakraborty:2011nf}. However, the introduction of temporal pulsations was also shown to negate this multi-angle matter suppression~\cite{Dasgupta:2015iia}. 

\begin{figure}[!t]
    \includegraphics[width=0.5\linewidth]{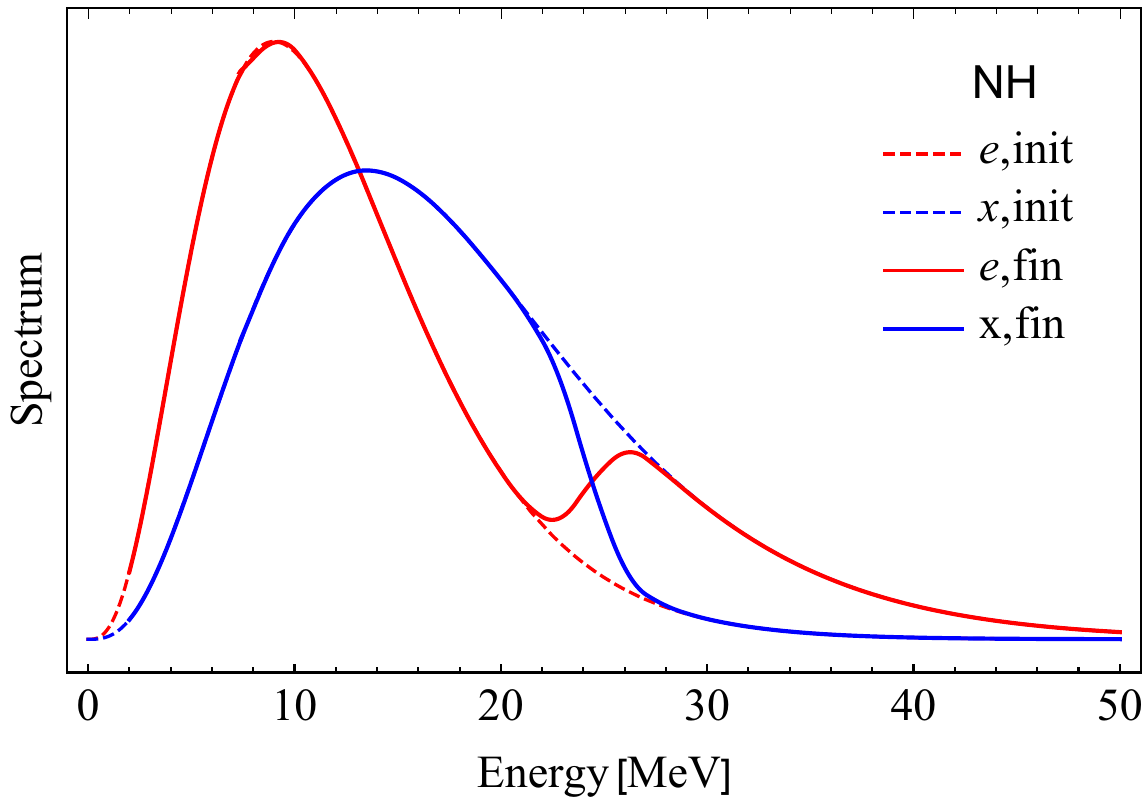}~~\includegraphics[width=0.5\linewidth]{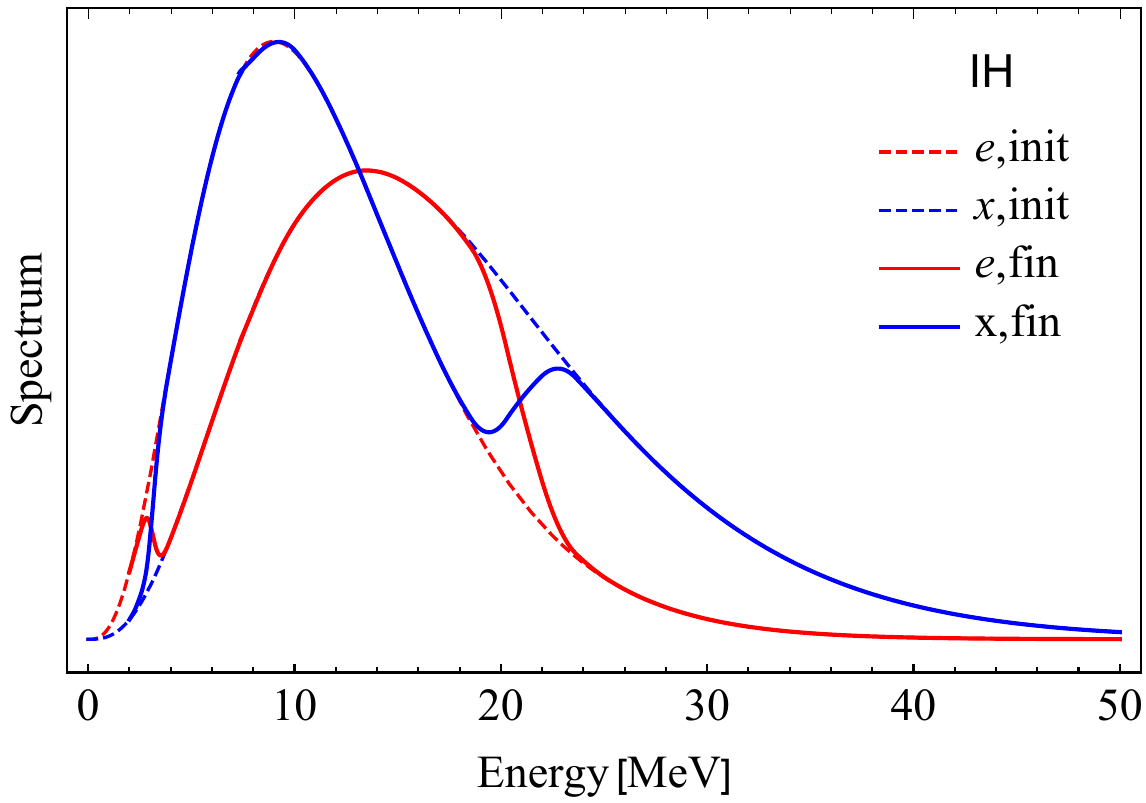}
    \caption{Spectral swap in the two mass orderings. Here ``$e$'' refers to $\nu_e$ whereas ``$x$'' refers to $\nu_x$.}
    \label{fig:split}
\end{figure}

While bipolar oscillations are periodic, in a real SN the neutrino number density decreases adiabatically. This was shown to lead to a spectral split,  where in IMO, all $\nu_e$ and $\nu_x$ within a certain energy range are swapped, whereas in NMO, the swapping occurs beyond a certain energy~\cite{Duan:2006an} (see Fig.\,\ref{fig:split}).
Such a flavour exchange is called a ``swap'', whereas the sharp boundaries at either side of the swaps are called ``splits''. A simple analytical understanding of the swaps was offered in terms of the neutrino spectra defined using the variable $\omega$~\cite{Raffelt:2007xt,Raffelt:2007cb}. Later works revealed the presence of multiple spectral splits in the cooling phase spectra~\cite{Fogli:2007bk,Fogli:2008pt}, which were explanined analytically in terms of 
of the development of spectral swaps in the difference spectrum $g_\w $, defined as~\cite{Dasgupta:2009mg} 
\begin{eqnarray}
\label{eq:diffspectrum}
g_\w &\propto& f_{\nu_e}(\w)-f_{\nu_x}(\w) \qquad{\rm for~ }\w>0 \; ,\nonumber\\
     &\propto& f_{\bar{\nu}_x}(\w)-f_{\bar{\nu}_e}(\w)\qquad {\rm for~}\w<0\,.
\end{eqnarray}
The development of a spectral swap is related to the presence of ``spectral crossings''. A crossing is defined as ``positive'' (``negative'') if $g_\w$  changes sign from negative (positive) to positive (negative). The emergence of spectral swaps around a crossing is intricately linked to the conservation of flavour lepton number during bipolar oscillations~\cite{Dasgupta:2009mg}. According to the equations of motion (EoMs), the quantity $\int d\omega g_\omega$ remains conserved, hence attaining a value of zero across the spectral swap. Consequently, spectral swaps can only materialise in proximity to a spectral crossing, which is also a necessary and sufficient condition for the flavor pendulum to be unstable.
Additionally, three-flavour effects were shown to induce additional splits, which could largely be understood as a sequence of two-flavour splits~\cite{Dasgupta:2008cd,Dasgupta:2007ws,Friedland:2010sc,Choubey:2010up}.

It quickly becomes evident that solving this complicated system of non-linear equations necessitates recourse to numerical simulations. However, gaining an intuitive grasp of the inception of these oscillations can be facilitated by interpreting them as an instability within flavour space and subjecting them to a linearised stability analysis\cite{Sawyer:2008zs,Banerjee:2011fj}. In this case, the density matrix is first expanded in the weak interaction basis as
\begin{equation}
\varrho_{\mathbf{p,r}}(t)=\frac{f_{\nu_e}+f_{\nu_x}}{2}+\frac{f_{\nu_e}-f_{\nu_x}}{2}\,
\begin{pmatrix}s_{\mathbf{p}}(t,\mathbf{r})&S_{\mathbf{p}}(t,\mathbf{r})\\S^*_{\mathbf{p}}(t,\mathbf{r})&-s_{\mathbf{p}}(t,\mathbf{r})\end{pmatrix}\,,
\end{equation}
where $f_{\nu_\alpha}$ represents the occupation number of $\nu_\alpha$. The real field $s_{\mathbf{p}}$ measures the evolution of the occupation number, whereas  $ S_{\mathbf{p}}$ measures the flavour coherence and hence is an indicator of the oscillation dynamics. Coherence dictates that $s^2 + |S|^2 =1$. 

Deep inside a SN, the matter density is high enough to assume that $s_{\mathbf{p}}\ll |S_{\mathbf{p}}|$
holds initially. As a result, Eq. \ref{eq:eom} can be expanded to linear order in $S_{\mathbf{p}}$. Using these linearised equations, one looks for plane wave solutions of the form $S_{\mathbf{p}}(t,\mathbf{r})=Q_\mathbf{p,k}\,e^{-i(k_0 t-\mathbf{k}\cdot\mathbf{r})}$, where $(k_0,\mathbf{k})$ denote the temporal and the spatial frequency of the flavour wave respectively. Using this in Eq. \ref{eq:eom} leads to a dispersion relation relating $k_0$ and $\mathbf{k}$, given by~\cite{Izaguirre:2016gsx,Capozzi:2017gqd} 
\begin{equation}
\rm{det}[\Pi^{\mu\nu}]=0\,,    
\end{equation}
where 
\begin{equation}\label{eq:Pi}
\Pi^{\mu\nu}=
\eta^{\mu\nu}+\int_{0}^{+\infty}\frac{E^2dE}{2\pi^2}\int \frac{d\mathbf{v}}{4\pi}\,g_{E,\mathbf{v}}\,
\frac{v^\mu v^\nu}{v_\alpha\,k^\alpha}\,.
\end{equation}
Here, $v^\mu=(1,\mathbf{v})$ denotes the velocity 4-vector of the neutrinos with $\mathbf{v}=\mathbf{p}/|\mathbf{p}|$, $k^\mu=(k^0,\mathbf{k})$, and $g_{E,\mathbf{v}}$ is the angle-dependent difference spectrum, similar to what is defined in Eq.\,\ref{eq:diffspectrum}.

If the dispersion relation is solved with an imaginary wave vector $\mathbf{k}$ for some real wave number $k_0$, the system exhibits spatial instability. Conversely, if the dispersion relation yields an imaginary temporal wave number $k_0$ for some real spatial wave vector $\mathbf{k}$, the system displays temporal instability. Although this method identifies whether a system is unstable, it does not provide insight into the extent of flavour conversions that may occur.

Instabilities are classified based on the growth rate of unstable modes, denoted by the imaginary part of the wave vector, ${\rm Im}(k)$. Instabilities with ${\rm Im}(k)$ proportional to $\sqrt{\omega \mu}$ are categorised as slow instabilities, while those with ${\rm Im}(k)$ proportional to $\mu$ are termed fast instabilities. Fast instabilities can further be classified as absolute or convective instabilities~\cite{Capozzi:2017gqd,Capozzi:2019lso,Yi:2019hrp}. Fast instabilities can reach ${\rm Im}(k) \sim O(1)$ cm$^{-1}$ in regions deep within the supernova core, where $\mu$ can be on the order of $O(10^5)$ km$^{-1}$. Consequently, fast instabilities are often considered to have probable impacts on the stellar dynamics, provided they are initiated deep insdie the SN. 

\subsection{Fast collective flavour conversions}
As discussed, flavour conversions are categorised as ``fast'' if they happen deep inside the supernova, at a distance of $\mathcal{O}({\rm cm})$ from the neutrinosphere. These rapid flavour conversions are mainly governed by the neutrino self-interaction strength $(\mu)$, and grow with the neutrino density. Therefore, they can take place even for $\Delta m^2\rightarrow 0$, requiring it only as numerical seeds, and hence are independent of the neutrino mass ordering. 
These solutions of the non-linear equations of motion were first discussed by Sawyer in a series of works~\cite{Sawyer:2005jk,Sawyer:2008zs,Sawyer:2015dsa}. This was later confirmed in~\cite{Chakraborty:2016lct,Dasgupta:2016dbv}, and this rekindled a wave of efforts in understanding the flavour propagation of neutrinos in dense media. 

\begin{figure}[!t]
    \includegraphics[width=0.7\linewidth]{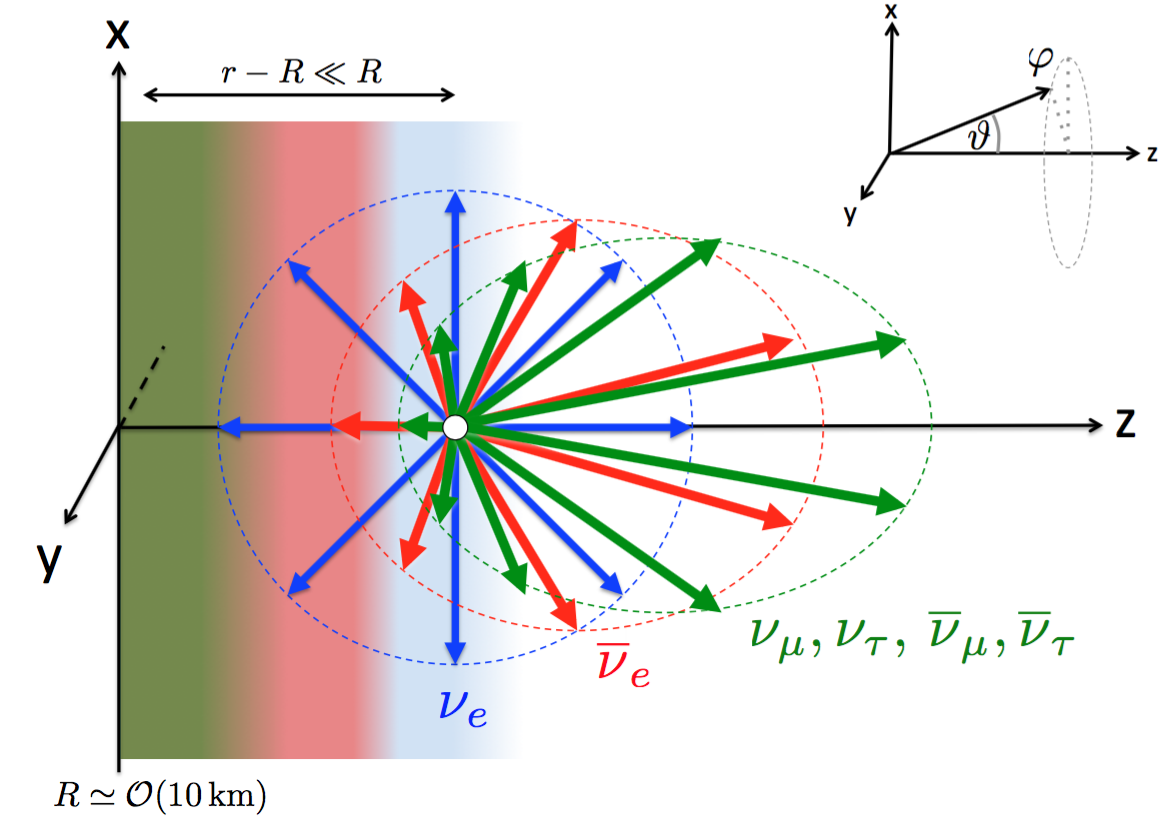}
    \caption{Schematic diagram of the neutrino angular distributions near the neutrinosphere, showing different flavours having different distributions. Neutrino flavours which decouple earlier have a more forward-peaked angular distribution, while those that decouple later are more isotropic. This leads to a zero-crossing in the ELN. Figure adapted from Ref.~\cite{Dasgupta:2016dbv}.}
    \label{fig:fastgeom}
\end{figure}

From the linear stability analysis, one can deduce that the condition required for the onset of the fast instability is that there exists at least one angular direction where the flux of $\bar{\nu}_e$ is greater than that of the $\nu_e$ (under the assumption that $\nu_{\mu,\tau}=\bar{\nu}_{\mu,\tau}$). This can also happen in scenarios where there is a backward flux of neutrinos. Hence, this is inherently a multi-angle phenomenon. The key to understanding the conditions for these fast conversions relies on the fact that different neutrino flavours start free-streaming at different radii, and hence near the neutrinosphere, they are expected to have different angular distributions (see Fig.\,\ref{fig:fastgeom}).

This can be formally defined by integrating Eq.\,\ref{eq:diffspectrum} to give the angular distribution of the flavour lepton number. Initial works in this direction defined the electron lepton number (ELN) as~\cite{Izaguirre:2016gsx}
\begin{equation}
    G_{\bf v}=\sqrt{2}G_F\,\int_{0}^\infty\frac{d E\, E^2}{2\pi^2} \left[f_{\nu_e}(E,{\bf v})-f_{\bar{\nu}_e}(E,{\bf v})  \right]\,
    \end{equation}
under the assumption that the non-electron flavours have the same angular distribution. In terms of this, it has been demonstrated, numerically as well as analytically, that a necessary and sufficient condition for the presence of fast modes, in an effective two-flavour setup, is the requirement of a zero-crossing in the ELN, i.e., the ELN changes sign for some angular modes~\cite{Abbar:2017pkh,Dasgupta:2021gfs,Morinaga:2021vmc}.

The assumption of equal angular distributions for the non-electron flavours was relaxed in~\cite{Chakraborty:2019wxe,Capozzi:2020kge}, which incorporated a stability analysis to study three flavour effects. This led to the definition of a corresponding muon lepton number $(\mu{\rm LN})$ and tau lepton number $(\tau{\rm LN})$, demonstrating that fast modes essentially depend on the zero-crossing between any two flavour lepton number, and not just the ELN alone. As a result, a positive zero-crossing in the  ELN can be removed by another positive zero crossing in the $\mu{\rm LN}$ or the $\tau{\rm LN}$. Therefore, fast modes can have very different behaviour in the three-flavour setup, as compared to a two-flavour setup. Follow-up works confirmed that this difference persists not only at the linear level but also at the non-linear regime, thereby emphasizing the necessity of a three-flavour study~\cite{Shalgar:2021wlj,Capozzi:2022dtr,Shalgar:2020xns}.    

The discovery of fast flavour conversions deep inside a SN-like toy setup immediately led to a dilemma - if flavour conversions indeed exist deep inside the SN medium, is it still valid to decouple neutrino inelastic collisions from flavour oscillations? It is a well-known fact that collisions can cause scattering-induced decoherence leading to damping of flavour oscillations.  This was first investigated using a simple two-neutrino beam model in 1+1 dimension~\cite{Capozzi:2018clo} and it was shown that although collisions are necessary to create the angular crossings required for fast oscillations, they are not strong enough to damp the flavour conversions. This was followed by a series of works, where the importance of the interplay of neutrino flavour conversions, inelastic collisions and advection was emphasised~\cite{Shalgar:2020wcx,Martin:2021xyl,Sasaki:2021zld,Sigl:2021tmj,Shalgar:2022lvv, Hansen:2022xza,Padilla-Gay:2022wck,Akaho:2023brj, Fiorillo:2023ajs,Shalgar:2023aca}.
More recently, a new class of instabilities, known as ``collisional instabilities'', has been found, where inelastic collisions are found to cause flavour instabilities~\cite{Johns:2021qby, Johns:2022yqy,Xiong:2022vsy,Lin:2022dek,Padilla-Gay:2022wck,Xiong:2022zqz,Liu:2023vtz,Kato:2023cig}. Inelastic collisions can trigger instabilities if the collisional rate changes with energy. Since neutrinos interact more than antineutrinos, this condition seems to be fulfilled in supernova. On the other hand, collisional instabilities have a much lower rate than fast instabilities (of the order of the collisional rate itself), so unless they are triggered in a region where fast instabilities are not, they would generally play a subdominant role; whether this can happen is an open question.
While considerable progress has been made in the study of the impact of inelastic collisions on fast flavour conversions, the final verdict is yet to be delivered.

\begin{figure}[!t]
    \includegraphics[width=0.7\linewidth]{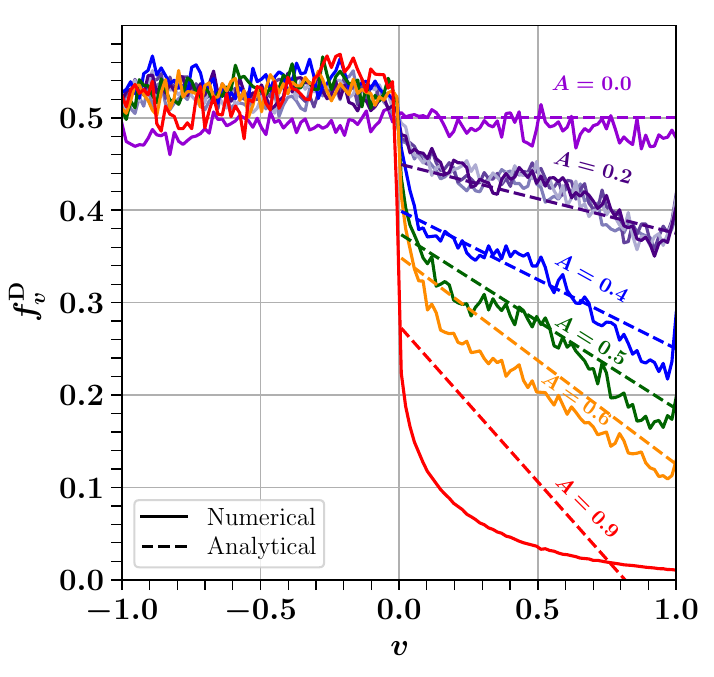}
    \caption{The depolarisation factor $f^D_v$ as a function of ${\mathbf v}$. It varies between 0 (no-depolarisation) and 0.5 (complete depolarisation), whereas $0.5\leq f^D_v \leq 1$ indicates a net flavour conversion. Figure adapted from Ref.\,\cite{Bhattacharyya:2020jpj}.}
    \label{fig:dep}
\end{figure}

The other major question eluding SN neutrino enthusiasts is regarding the outcome of fast conversions and its impact on SN neutrino spectra. This requires going beyond a linearised analysis and studying the solutions to the equations of motion in the non-linear region~\cite{Sawyer:2008zs}. This problem is hard to solve analytically and requires various simplifications to proceed. An attempt at an analytical solution in the non-linear regime was developed in~\cite{Dasgupta:2017oko}, where it was shown that a system of four intersecting neutrino beams undergoing fast oscillations resembles a particle oscillating in a quartic potential well.
Ref.~\cite{Abbar:2018beu} studied the evolution of fast modes in the non-linear regime numerically 
and deduced that it is not necessary for the final survival probability to approach a complete quasi-steady flavour equilibration. Ref.~\cite{Bhattacharyya:2020dhu,Bhattacharyya:2020jpj} demonstrated through analytical as well as numerical techniques that at late times, a system undergoing fast oscillations reaches a steady state and the most likely outcome is a rapid mixture of all flavours, leading to \emph{fast flavour depolarization} with the conservation of flavour lepton number (see Fig.\,\ref{fig:dep} which depicts the depolarisation factor $f^D_v$ as a function of ${\mathbf v}$). This was subsequently confirmed in~\cite{Wu:2021uvt,Richers:2021nbx,Richers:2021xtf,Bhattacharyya:2022eed,Zaizen:2022cik}. This has been further investigated recently in Ref.\,\cite{Cornelius:2023eop}, which studied the impact of imposing boundary conditions on the final fate of fast flavour conversions and demonstrated that flavour equipartition can be absent if the underlying boundary condition in a given simulation setup is not periodic. A number of other papers have also explored the non-linear regime of fast flavour conversions of neutrinos to understand the final fate of these flavour conversions~\cite{Johns:2020qsk,Sigl:2021tmj,Xiong:2021dex,Sasaki:2021zld,Abbar:2021lmm,Richers:2022bkd,Nagakura:2022kic,Zaizen:2023ihz, Fiorillo:2023hlk, Abbar:2023ltx, Abbar:2024chh, Fiorillo:2024fnl, Fiorillo:2024qbl,Xiong:2024pue}.
Finally, studies have also developed the analogy with that of a flavour pendulum, much like in the bipolar (slow-collective) oscillations case~\cite{Johns:2019izj,Padilla-Gay:2021haz,Bhattacharyya:2022eed,Fiorillo:2023hlk,Fiorillo:2023mze}. In this case, the final fate of fast flavour conversions was shown to be related to the relaxation of the flavour pendulum. Note that, however, these analogies are strictly valid for purely homogeneous systems. The presence of any inhomogeneity can get exponentially  enhanced due to fast instabilities. In such a case, the final outcome is expected to be a depolarized state, with flavour changes on very small scales.

The potential impact of fast flavour conversions on supernova dynamics prompted numerous investigations into the occurrence of rapid instabilities in various SN models. An initial exploration in this direction was undertaken in~\cite{Tamborra:2017ubu}, where a specialised examination of the angular distributions of neutrino radiation fields in several spherically symmetric (1D) SN simulations failed to identify any crossings in the electron lepton number (ELN) near the neutrinosphere. However, the presence of a large-scale dipole asymmetry in the ELN emission, described as Lepton-Emission Self-sustained Asymmetry (LESA)~\cite{Tamborra:2014aua}, has led to the possibility of the presence of an ELN crossing in multi-dimensional SN simulations. This has motivated numerous studies in the search for the conditions of fast conversions in state-of-the-art, self-consistent multidimensional SN models~\cite{Dasgupta:2018ulw,Glas:2019ijo,Nagakura:2021txn,Abbar:2020fcl,Abbar:2020qpi,Capozzi:2020syn,Nagakura:2021suv,Nagakura:2021hyb,Ehring:2023lcd,Abbar:2023zkm,Froustey:2023skf}. Most of the works rely on the fact that SN simulations generally do not provide the full angular distributions of neutrino flavours, but the moments of the angular distributions. Hence the studies focus on optimizing the use of these angular moments to look for conditions conducive to the presence of fast conversions.  

\begin{figure}[!t]
    \includegraphics[width=0.8\linewidth]{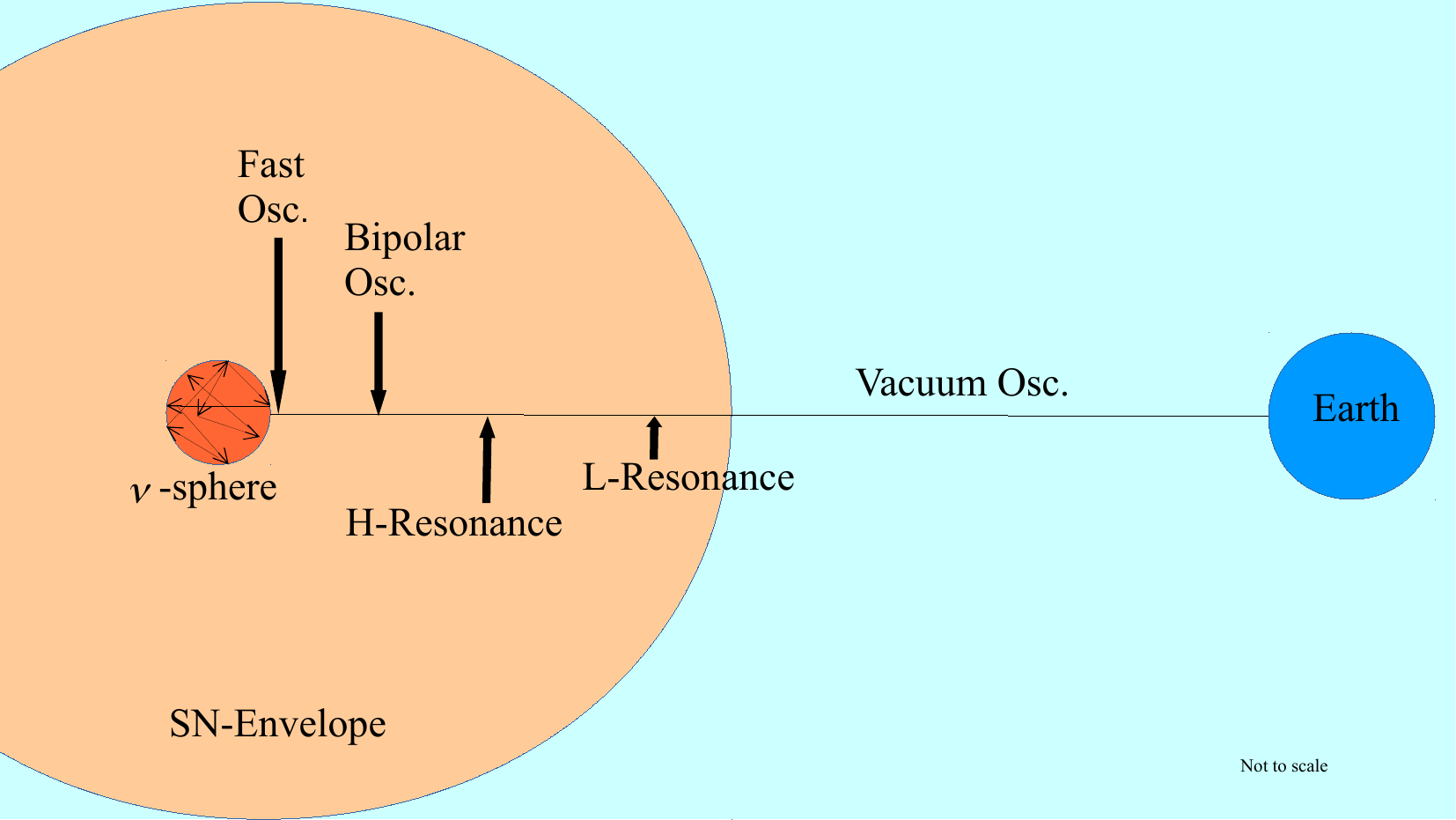}
    \caption{Schematic diagram denoting different regions of flavour conversion inside a SN envelope. Figure adapted from Ref.\,\cite{Sen:2018put}.}
    \label{fig:SNGeom}
\end{figure}

Flavour conversions of neutrinos deep inside a SN represent a significant area of uncertainty in a comprehensive, first-principle modelling of the SN dynamics. Our studies over the past decades have unearthed different types of flavour conversion mechanisms operating at different distances from the core, as depicted in Fig.\,\ref{fig:SNGeom}.
Significant progress is being made on the impact of flavour conversions on the explosion dynamics- depending on the extent of flavour conversions, explosion could be either hindered or assisted~\cite{Dasgupta:2011jf, Ehring:2023lcd,Ehring:2023abs} as well as the  nucleosynthesis of heavy elements~\cite{Fujimoto:2022njj,Xiong:2020ntn, Fischer:2023ebq,Balantekin:2023ayx,Friedland:2023kqp}. However, we are still far from a complete picture and the final verdict on the outcome is yet to be out.

\subsection{A comment on mean-field approximation \emph{vs.} many-body effects}
Discussions on the dynamics of collective oscillations have primarily relied on a mean-field approximation, neglecting all quantum correlations among neutrinos. Despite the wide use of this approach, its validity has been a matter of debate over the past years \cite{Friedland:2003eh,Friedland:2003dv,Balantekin:2006tg,Pehlivan:2011hp,Volpe:2013uxl, Vaananen:2013qja,Serreau:2014cfa}.
Hence, it is crucial to assess the applicability of this approximation across various scenarios to ensure the accurate interpretation of results and to delve into the concept of entanglement among the neutrinos within the many-body system. Over the last few years, there has been growing effort in this direction, with the aim of understanding the efficacy of a mean-field treatment of neutrinos in such a dense environment~\cite{Birol:2018qhx,Patwardhan:2019zta,Cervia:2019res,Rrapaj:2019pxz,Roggero:2021asb,Patwardhan:2021rej}. We will not go into the details of this discussion, but refer the interested readers to~\cite{Patwardhan:2022mxg} (and references therein) for more detailed review.

\section{New physics from supernova neutrinos}
\label{sec:newphys}
A galactic core-collapse SN can act as the ultimate particle physics laboratory to probe physics beyond the Standard Model~\cite{Raffelt:1996wa,Raffelt:1999tx}. The extreme environment allows for probing a wide range of new physics, which cannot otherwise be accessible in terrestrial laboratories. From a phenomenological point of view, new physics inside a supernova can have either of the following effects:
\begin{enumerate}
    \item It can affect the net luminosity and average energy of the emitted neutrinos, and hence the duration of the neutrino emission phase.
    \item It can affect the shape/normalisation of the emitted neutrino spectra.
\end{enumerate}
Using a combination of one or both of these effects, strong constraints can be imposed on new physics from a galactic supernova.

The most well-known of the bounds coming from SN1987A is the anomalous cooling bound~\cite{Ellis:1987pk,Raffelt:1987yt,PhysRevLett.60.1797,Burrows:1990pk, Raffelt:1996wa}. Neutrinos from SN1987A can be used to constrain the amount of energy that can be transported by any new degree of freedom, weakly coupled to the SM. Such particles, if produced inside a supernova, can lead to new modes of energy loss and hence additional cooling channels. Based on this, a criterion for the maximum luminosity of the new particle was proposed. If the luminosity of the new particle, $\mathcal{L}\gtrsim 3\times 10^{52}\,{\rm erg \,s}^{-1}$ for $\rho=3\times 10^{14}\,{\rm g/cc}$ and $T=30\,{\rm MeV}$, then the duration of the neutrino burst is reduced from the observed $\sim 10\,{\rm s}$ of SN1987A. This argument leads to a bound on the coupling of the new particles. However, the bound cannot be applied for arbitrarily large couplings. For couplings stronger than a certain value, the new particles get trapped and cannot escape the SN core, leading to an exclusion window in the coupling-mass parameter space. On the other hand, for couplings which are too weak, the particles may not be efficiently produced to alter the dynamics of a SN.  

This argument has been used to constrain new weakly coupled particles like the axion~\cite{Raffelt:1987yt,Dolan:2017osp, Chang:2018rso,Carenza:2019pxu,Carenza:2020cis,Lucente:2020whw,Calore:2021klc,Caputo:2024oqc}, dark gauge bosons charged under new symmetries~\cite{Rrapaj:2015wgs,Chang:2016ntp,Sung:2021swd,Caputo:2021eaa,Cerdeno:2023kqo}, majorons and other light scalars coupled to neutrinos~\cite{Farzan:2002wx,Brune:2018sab, Heurtier:2016otg, Diamond:2023scc,Antel:2023hkf}, and so on. More recently, the cooling argument was refined to include physical effects which were ignored previously and a detailed prescription was laid down for limits from more updated models of SNe~\cite{Caputo:2021rux,Caputo:2022rca}. However, note that in cases where there exists significant self-interactions in the dark sectors, these cooling arguments may not apply~\cite{Fiorillo:2024upk}. Modern SN codes nowadays can perform a complete 6-flavour neutrino (antineutrino) transport, which has demonstrated the relevance of muons in simulations~\cite{Bollig:2017lki}. This has led to additional constraints on \emph{muon-philic} new degrees of freedom~\cite{Bollig:2020xdr, Croon:2020lrf,Caputo:2021rux}. The dynamics of a core-collapse can also be altered by the decay of bosonic particles into neutrinos~\cite{Fiorillo:2022cdq}, or photons~\cite{Caputo:2022mah} deep inside the SN progenitor.

Similar cooling arguments have also been put forth to constrain the mass and mixing of additional species of sterile neutrinos $(\nu_s)$. It is well known that sterile neutrinos with masses $\sim \mathcal{O}(100\,{\rm MeV})$ can be produced from different processes inside the SN core~\cite{Mastrototaro:2019vug,Carenza:2023old,Brdar:2023tmi}. These sterile neutrinos, if produced, can stream out of the SN, thereby leading to additional cooling, and hence can be constrained.
Lower mass (keV) sterile neutrinos  are excellent dark matter candidates. The $\nu_s$  can be easily produced inside a SN through adiabatic flavour conversion at MSW resonances, or through a scattering-induced decoherence~\cite{Raffelt:2011nc, Arguelles:2016uwb,Syvolap:2019dat,Suliga:2019bsq,Chen:2022kal,Suliga:2020vpz}.  The usual cooling arguments apply here as well.  However, since the $\nu_s$ is produced from an active neutrino, this in turn affects the forward scattering potential that the propagating neutrinos experience. This leads to feedback, which when taken into account consistently can loosen the bounds on the sterile mass-mixing parameter space~\cite{Suliga:2019bsq,Suliga:2020vpz}. However, secret interactions of neutrinos can spoil the party, and enhance the production of $\nu_s$, which can potentially overcome this feedback effect~\cite{Chen:2022kal}. Lighter mass $\nu_s$ (eV scale) can be produced from active neutrinos directly via resonant MSW transitions at densities that can affect the heating of low mass SNe and affect their explosion~\cite{Tamborra:2011is,Franarin:2017jnd,Tang:2020pkp}. For larger mass SNe, the production happens at densities which can affect the active neutrino flux, and hence have important consequences for nucleosynthesis~\cite{Tamborra:2011is}. 

Interactions of neutrinos beyond the SM can also be probed effectively inside a SN. Non-standard interactions (NSI) of neutrinos with matter can lead to additional MSW resonances inside the SN (like the I-resonance~\cite{Esteban-Pretel:2007zkv,Esteban-Pretel:2009jqw}), thereby altering the neutrino flavour evolution. These interactions can change the neutron-proton ratio, leading to intriguing consequences for nucleosynthesis yields. Similarly, a core-collapse SN is one of the few laboratories where non-standard neutrino self-interactions (NSSI) can be tested~\cite{Blennow:2008er,Das:2017iuj,Dighe:2017sur}. In fact, a smoking-gun signature of the presence of flavour-violating NSSI will be the presence of spectral splits in the neutronisation burst spectra due to collective oscillations, which do not take place within the SM in this time phase~\cite{Das:2017iuj}. Furthermore, the presence of neutrinophilic bosons can lead to $2\nu \rightarrow 4\nu$ processes inside the SN, which reduces the net energy deposited behind the shock, thereby hindering explosion~\cite{Shalgar:2019rqe}, however the inclusion of the inverse reaction can counter this effect~\cite{Fiorillo:2023ytr}. 
Secret interactions of neutrino have also been argued to extend the duration of the SN neutrino burst~\cite{Chang:2022aas}, but these claims were recently countered in~\cite{Fiorillo:2023ytr}. Secret interactions of neutrinos can also be constrained from their interactions with the cosmic neutrino background, which leads to a dissipation of the neutrino energy below the expected value~\cite{Shalgar:2019rqe}.
 Neutrinos from a SN can also be used to probe neutrino-dark matter interactions through their interaction with the DM in the galactic halo~\cite{Carpio:2022sml,Das:2021lcr,Lin:2022dbl,Das:2024ghw}.

The other direction where SN neutrinos can contribute immensely is towards constraining  neutrino parameters. The large neutrino flux, combined with the propagation through extreme conditions, allows one to put stringent constraints on neutrino properties. For example, using SN1987A data, time of flight constraints were used to set neutrino mass bounds to $m_\nu<20\,{\rm eV}$~\cite{Kernan:1994kt}. Future experiments like DUNE can use the neutronization burst phase to narrow it down to $\mathcal{O}(1)\,{\rm eV}$~\cite{Pompa:2022cxc}. Of course, this requires a good estimation to the distance of the next galactic SN, which can be achieved via triangulation~\cite{Hansen:2019giq,Linzer:2019swe}.

The neutronization burst is an excellent tool to probe the neutrino mass-ordering, which is one of the major unknowns in neutrino physics. Deep inside the supernova, neutrinos are generated within regions of exceedingly high local matter densities. Conventional matter effects dictate that the $\nu_e$ from the burst phase align closely with the instantaneous Hamiltonian eigenstate associated with the highest instantaneous Hamiltonian eigenvalue upon production. Subsequent evolution treats the distinct eigenstates as incoherent, with the adiabatic approximation proving effective owing to the spatial variation of matter density and known neutrino mass-squared differences. Consequently, the neutronization-burst neutrino flux at Earth, when disregarding the minor initial $\nu_x$ population, can be well approximated by \cite{Dighe:1999bi}:
\begin{equation}
f_{\nu_e}(E,t) = \frac{1}{4\pi R^2}|{\rm U_{eh}}|^2 \Phi_{\nu_e}(E,t)
\label{eq:NueEarth}
\end{equation}
where $h$ denotes the heaviest neutrino mass eigenstate, with $h=3$ for the NMO and $h=2$ for the IMO, ${\rm U_{eh}}$ represents the relevant element of the leptonic mixing-matrix, while $R$ signifies the distance between Earth and the supernova.  Given the negligible production of other flavours during this phase, the $\nu_e$ flux for the NMO experiences a relative suppression factor of approximately $\sim |{\rm U_{e3}}|^2/ |{\rm U_{e2}}|^2 \simeq 0.1$. This facilitates the determination of the neutrino mass ordering during the neutronization-burst phase \cite{Dighe:1999bi}. The mass-ordering can also be deduced from timing information in upcoming detectors~\cite{Serpico:2011ir,Brdar:2022vfr}.

Finite neutrino masses imply that neutrinos can decay. Neutrinos from a SN can be used to put tight constraints on invisible, visible as well as radiative
decays of neutrinos~\cite{Raffelt:1999tx}. While data from SN1987A constrains the invisible decays to $\tau/m > 10^5\,{\rm s/eV}$~\cite{Frieman:1987as, Ivanez-Ballesteros:2023lqa} , future galactic SN can narrow it down to $\tau/m \lesssim  10^6\,{\rm s/eV}$ for a SN happening at 10\,kpc~\cite{deGouvea:2019goq}. Radiative decays of neutrinos can lead to a coincident $\gamma$ flare - this has been used to constrain radiative decays to $\tau/m \gtrsim 10^{15}\,{\rm s/eV}$~\cite{Raffelt:1999tx}. 

The strong magnetic field existing inside a SN can also be used to probe neutrino electromagnetic properties. For Dirac neutrinos, a non-zero intrinsic magnetic moment can give rise to a spin flip for a certain flavour $\alpha$, i.e. $\nu_{\alpha L} \rightarrow \nu_{\alpha R}$, while a transition magnetic moment also flips the  flavour, i.e., $\nu_{\alpha L} \rightarrow \nu_{\beta R}$~\cite{Pal:1981rm}. This conversion process, $\nu_L\rightarrow \nu_R$, leads to faster cooling of the SN, and can be used to constrain the magnetic moment $\mu_\nu \leq 10^{-11}\, \mu_B$~\cite{Barbieri:1988nh}.
Furthermore, in the presence of a magnetic field, coherent scattering of neutrinos off the background can resonantly enhance these spin-flavour transitions, known as resonant spin-flavour precession (RSFP)~\cite{ Lim:1987tk, Akhmedov:1988uk}. Such RSFP can lead to intriguing signatures in the neutronisation spectra~\cite{ Ando:2003is, Akhmedov:2003fu} and can be used to constrain non-zero transverse magnetic moments much better than current laboratory constraints~\cite{Jana:2022tsa}. Neutrino magnetic moments have also been shown to be relevant for collective oscillations ~\cite{deGouvea:2012hg,Abbar:2020ggq}. One can also constrain the effective neutrino electric charge $(q_\nu)$ from the observation of neutrinos from SN1987A. The presence of galactic and extragalactic magnetic fields may elongate the trajectory of millicharged neutrinos, thereby delaying their arrival time at the Earth beyond the observed period of a few seconds. This constrains $q_{\nu_e} < 2\times 10^{-15}\,e$ for an intergalactic field $B=10^{-3}\,\mu G$ and a distance of $R=50\,{\rm kpc}$, and  $q_{\nu_e} < 2\times 10^{-17}\,e$ for a galactic field  $B=1\,\mu G$ and a SN at a distance of $R=10\,{\rm kpc}$~\cite{Giunti:2014ixa}.

The nature of neutrinos is another mystery that has not been solved yet. Depending on whether lepton number is conserved or not, neutrinos can be Dirac or Majorana. Supernova neutrinos can also be used to pin down the nature of neutrinos, \emph{in the presence of new physics}.  For example, studies have shown that for a future galactic SN, using a combination of upcoming experiments like DUNE and HK, one can distinguish between a decaying Dirac and a decaying Majorana neutrino from the neutronization burst flux~\cite{deGouvea:2019goq}. Similarly, in the presence of non-zero magnetic moments, Dirac and Majorana can have different signatures in the neutrino flux~\cite{Jana:2022tsa}. 
However, there also exists the possibility of soft lepton number violation (LNV). The degree of LNV can be gauged by the relative smallness of the Majorana mass term compared to the Dirac mass term for neutrinos. In such a scenario, neutrinos exhibit pseudo-Dirac (or quasi-Dirac) behaviour~\cite{Wolfenstein:1981kw}. The soft nature of LNV ensures that although neutrinos are Majorana, they effectively behave akin to Dirac neutrinos in practical terms. In this scenario, active-sterile neutrino oscillations are driven by a minute mass-squared difference, denoted as $\delta m^2$, between the mass eigenstates. However, this mass-squared difference is only accessible over astronomically large baselines. Strong constraints can be set on this mass-squared difference from the neutrino events from SN1987A ~$\delta m^2\lesssim 10^{-20}~{\rm eV}^2$~\cite{Martinez-Soler:2021unz, Sen:2022mun}.

This is just an incomplete list of the different kinds of particle physics constraints one can probe from the observation of supernova neutrinos. Studies have shown that the few neutrino events from SN1987A can also be used to probe the deviations of the equivalence principle~\cite{Guzzo:2001vn}, the gravitational memory effect~\cite{Mukhopadhyay:2020ubs}, the effect of neutrino long-range forces~\cite{Reddy:2021rln,Ghosh:2022nzo}, consequences of deviation from Lorentz symmetry in the neutrino sector~\cite{Chakraborty:2012gb}, the QCD phase transition~\cite{Dasgupta:2009yj}, and several other exotic ideas which are otherwise difficult to test. This is a testament to the immense potential offered by a future galactic SN to test physics beyond the SM. In conclusion, we would also like to point to the potential of the diffuse supernova neutrino background - a sea of neutrinos from all possible core-collapse supernovae (and failed supernovae) in the Universe - to probe fundamental physics (see \cite{Beacom:2010kk, DeGouvea:2020ang, Ando:2023fcc} and references therein for a detailed discussion of this aspect).

\section{Conclusion}
\label{sec:conc}
The study of neutrinos from supernovae represents a captivating and multifaceted field of research requiring significant contributions from particle physics and astrophysics. Through decades of theoretical modelling, and computational efforts, our understanding of the intricate processes governing neutrino emission, propagation, and interaction within the surroundings has advanced significantly. We have strong reasons to believe that neutrinos play a crucial role in the supernova explosion mechanism, as well as in the formation of heavy elements inside stars. Furthermore, these neutrinos offer the best bet to probe the physics of dense matter existing inside a supernova, which cannot be achieved inside a terrestrial laboratory. As a result, a careful study of these neutrinos is necessary to utilise the full potential of the next galactic supernova.

In this brief review, we discussed the major advancements made in the field and where we stand. While we have made remarkable progress in the understanding of neutrino flavour oscillations in dense environments, we still do not know the exact impact of these effects. The numerical complexity associated with the complete problem necessitates the introduction of simplifications, which can sometimes hide additional flavour instabilities. Previously, the relaxation of such symmetries has led to the discovery of new instabilities associated with neutrino non-radial evolution inside a supernova. A recent example is that of fast pairwise flavour conversions, which was missed due to the assumption of equal angular distributions for neutrinos and antineutrinos. These rapid flavour conversions, occurring deep inside a supernova, are almost independent of the neutrino mass and mixing, and hence unsuppressed due to matter effects. Recent studies indicate that these fast oscillations eventually lead to an almost flavour equilibration deep inside the supernova, however further studies are required to confirm this idea. As a result, there has been a surge in research efforts aimed at evaluating the impact of these flavour conversions on the mechanism of supernova explosions.

We also highlighted how the neutrino flux from a supernova allows for imposing stringent constraints on neutrino properties, which are typically unattainable in terrestrial laboratories. This advocates a core-collapse supernova as the ultimate laboratory for particle physics. Therefore, the study of supernova neutrinos remains a vibrant and fruitful avenue of exploration, offering invaluable insights into the fundamental nature of matter and the dynamics of stellar evolution.

\acknowledgments
We would like to thank Basudeb Dasgupta, Georg Raffelt and Irene Tamborra for a careful reading of the manuscript and insightful comments. We also acknowledge the hospitality of the Network for Neutrinos, Nuclear Astrophysics, and Symmetries (N3AS) Physics Frontier Center of UC Berkeley where part of this review was completed.

\bibliography{references.bib}

\end{document}